\documentclass[%
 reprint,
superscriptaddress,
 amsmath,amssymb,
 aps,
prl,
]{revtex4-2}
\usepackage{graphicx}
\usepackage{dcolumn}
\usepackage{bm}
\usepackage{braket}

\begin{document}

\preprint{APS/123-QED}

\title{Self-stabilization of microcombs}

\author{Krishna Twayana}
 \affiliation{Department of Microtechnology and Nanoscience, Chalmers University of Technology SE-41296 Gothenburg, Sweden }
 
\author{Fuchuan Lei}
 \affiliation{State Key Laboratory of Integrated Optoelectronics, Key Laboratory for UV-Emitting Materials and Technology of Ministry of Education, School of Physics, Northeast Normal University, Changchun 130024, China}

\author{Junyong Choi}%
\affiliation{School of Mechanical and Aerospace Engineering, Korea Advanced Institute of Science and Technology (KAIST), Daejeon 34141, Republic of Korea
}%

\author{Jungwon Kim}%
\affiliation{School of Mechanical and Aerospace Engineering, Korea Advanced Institute of Science and Technology (KAIST), Daejeon 34141, Republic of Korea
}%

\author{Victor Torres-Company}%
 \email{torresv@chalmers.se}
\affiliation{Department of Microtechnology and Nanoscience, Chalmers University of Technology SE-41296 Gothenburg, Sweden
}%

\date{\today}

\begin{abstract}
Optical frequency combs form phase-locked spectral lines arranged on an equidistant grid fully defined by two degrees of freedom, i.e., the repetition rate and frequency offset. Stabilizing these parameters to a common frequency reference results in a coherent frequency ruler, central for modern precision metrology. However, extending this level of stability to chip-scale microcombs remains an outstanding challenge. Here, we demonstrate a self-stabilizing mechanism based on self-injection locking of a selected comb line via an external feedback loop. This process establishes a second anchor point in addition to the pump, thereby constraining the comb’s frequency noise dynamics. We show that, with an appropriate choice between pump frequency noise and feedback strength, collective fluctuations of the repetition rate are strongly suppressed. The result is a microcomb exhibiting ultralow phase noise and dramatically reduced timing jitter. In a 100 GHz Si\textsubscript{3}N\textsubscript{4} soliton microcomb, we achieve an unprecedented combination of high-conversion efficiency, sub-Hertz intrinsic linewidth across the entire C band, and an integrated timing jitter of 1 fs. This approach enables chip-scale microcombs with remarkable noise performance and fs-level pulse stability, surpassing conventional noise limits and opening new avenues for precision metrology at the chip scale.
\end{abstract}

\maketitle
The optical spectrum of a periodic pulse train, such as that generated by a mode-locked laser, consists of a set of evenly spaced frequency lines arranged on an equidistant grid. Because the optical carrier frequency of the pulses is generally not an integer multiple of the repetition rate, the grid is fully defined by two independent parameters: the repetition rate (line spacing) and the carrier-envelope-offset frequency \cite{udem2002optical}. Stabilizing these two degrees of freedom to an atomic frequency reference transforms the spectrum into a fully stabilized optical frequency comb \cite{udem1999absolute,diddams2000direct}. Such combs have become indispensable tools in precision metrology, enabling advances ranging from optical atomic clocks \cite{diddams2001optical, wu2025vernier, fortier2026optical} to ultralow-noise microwave synthesis \cite{xie2017photonic, fortier2011generation, zhao2024all}.

\begin{figure*}[tb]
\centering
\includegraphics[width=\linewidth]{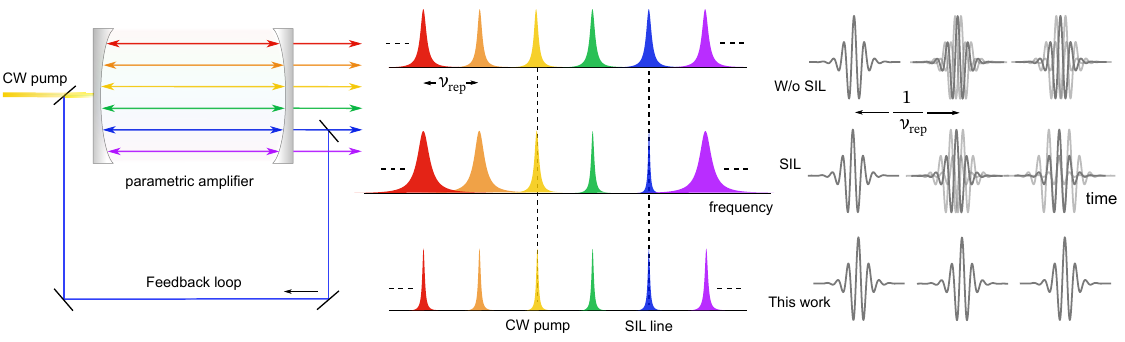}\caption{Self-injection locking parametric amplifier in a cavity system (left). Optical parametric oscillators mode profile (center) and timing jitter (right) without and with optical feedback.}
\label{fig:Fig1} 
\end{figure*}

In the free-running regime, when the repetition rate and offset frequency are not stabilized, the frequency grid becomes dynamic, with its performance governed by the intrinsic noise processes of the optical frequency comb \cite{lei2022optical, jin2024self, bao2021quantum}. Fluctuations of the two defining parameters collectively broaden the comb lines and produce a mode-number-dependent frequency instability, commonly described as the elastic tape model \cite{telle2002kerr}. Remarkably, this model is universal, i.e., it arises directly from the stochastic dynamics of the two degrees of freedom defining the comb grid, independent of the specific laser architecture \cite{takushima2004linewidth, schmeissner2014spectral, liehl2019deterministic}. 
Direct stabilization of the repetition rate and offset frequency to a common frequency reference suppresses these collective fluctuations, narrowing the comb linewidths and transferring the stability of the external reference to the entire spectrum. This approach, however, relies on self-referencing, a demanding technique that remains challenging to implement in chip-scale platforms \cite{drake2019terahertz, brasch2017self} such as microcombs using fully integrated photonics. 
To circumvent these limitations, alternative techniques have been developed, such as two-point locking \cite{papp2014microresonator} and Kerr optical synchronization \cite{moille2023kerr}. These strategies pin the two degrees of freedom of the frequency grid to two independent and spectrally separated continuous-wave (CW) lasers \cite{swann2011microwave}. In doing so, they establish two anchor points for the frequency grid while relaxing the bandwidth requirements associated for self-referencing \cite{fortier2011generation}. These approaches have enabled ultrastable pulse trains and pure microwave signal generation \cite{kudelin2024photonic, sun2024integrated, jin2025microresonator, sun2025microcavity, ji2025dispersive}. 
A conceptually different strategy, first introduced in \cite{lei2024self}, derives the second anchor point from within the microcomb spectrum \cite{li2025dispersive}. In this scheme, a selected comb line is subject to optical feedback. When the feedback strength exceeds a critical level, that line becomes self-injection locked and undergoes a dramatic linewidth reduction relative to the free-running state. The locked sideband and the pump itself act as the two anchor points, in a manner akin to Kerr optical synchronization \cite{moille2023kerr}. Crucially, however, the injection-locked line typically attains a narrower linewidth than the pump. According to the elastic tape model, this imbalance redistributes noise across the spectrum, leading to increased frequency fluctuations for comb lines on the opposite side (Fig.\ref{fig:Fig1}). 

Here, we demonstrate both numerically and experimentally that the elastic-tape model remains valid in the sideband self-injection-locking regime. This insight reveals a route to strongly suppress repetition rate fluctuations by judiciously optimizing the feedback strength and pump linewidth. Implementing this approach in a silicon nitride microcomb based on coupled cavities \cite{helgason2023surpassing}, we reach sub-Hz intrinsic linewidth across the spectrum. Remarkably, the ultra-narrow-linewidth operation is encompassed by a dramatic reduction in timing jitter, reaching the femtosecond level.

\section{Elastic tape model under sideband self-injection locking}

In free running microcombs, the intrinsic noise leads to collective fluctuation of the comb lines, following the elastic tape model. Concretely, the intrinsic linewidth of the $\mu$ comb line counted away from the pump is
\begin{equation}    
\Delta\nu_\mu=\mu^2\Delta\nu+\Delta\nu_p
\label{equa1}
\end{equation}
This equation indicates that the frequency noise power spectral density of the $\mu$th comb line follows a symmetric quadratic distribution as the number of comb lines increases. The coefficient $\Delta\nu=\Delta\nu_\text{Q}+\Delta\nu_\text{TRN}+\Delta\nu_\text{RIN}$ incorporates contributions from the pump’s relative intensity noise, $\Delta\nu_\text{RIN}$, quantum timing jitter, $\Delta\nu_\text{Q}$, and thermorefractive noise, $\Delta\nu_\text{TRN}$. For simplicity, we consider that the pump’s phase noise is sufficiently low, so that the coupling between repetition rate and pump’s frequency fluctuations induced by the Raman self-frequency shift is negligible. These results have been analyzed in detail in reference \cite{lei2022optical}. 

Under sideband self-injection-locking, the comb lines near the self-injection-locking point experience a dramatic noise reduction, but the elastic-tape model still applies. The frequency mode experiencing minimum frequency fluctuations moves away from the pump, resulting in a redistribution of frequency noise, i.e., 
\begin{equation}
\Delta\nu'_\mu=(\mu-\mu_q)^2\Delta\nu'+\Delta\nu_{\mu_q}
\label{SIL:linewidth}   
\end{equation}
with $\Delta\nu_{\mu_q}$ as the quite point mode ($\mu_q$) linewidth. With two reference points given by the pump (0, $\Delta\nu_p$) and the self-injection-locked sideband ($\mu_\text{SIL}$, $\Delta\nu_\text{SIL}$), the quadratic coefficient ($\Delta\nu'$) is evaluated as
\begin{equation}
    \Delta\nu'=\frac{\Delta\nu_P-\Delta\nu_\text{SIL}}{\mu_\text{SIL}(2\mu_q-\mu_\text{SIL})}
\end{equation}
This equation indicates that if the pump frequency noise and feedback strength (which determines $\Delta\nu_\text{SIL}$) are chosen such as $\Delta\nu_P \approx \Delta\nu_\text{SIL}$, the comb lines in the spectrum will reach the lowest possible frequency noise, $\Delta\nu_{\mu_q}$, across the whole spectrum. Physically, when the feedback strength and pump linewidth are chosen such that the locked sideband and the pump attain comparable residual frequency noise, the stretch coordinate is efficiently damped without introducing a strong imbalance between the anchors. The grid then evolves predominantly through common-mode translation, so the frequency noise becomes nearly mode independent and all comb lines approach the same noise floor.

The architecture of self-injection locking (SIL) in a parametric oscillator with an external feedback loop is shown in Fig. \ref{fig:Fig1}. In this scheme, one of the generated cavity modes is fed back to the input, effectively extending its photon lifetime and reducing its phase noise and linewidth. While SIL reduces the frequency noise of the locked mode, it can increase the noise of other modes depending on the feedback ratio. The larger the difference  $|\Delta\nu_p-\Delta\nu_\text{SIL}|$, the steeper the resulting linewidth distribution, which leads to increased timing jitter compared to the free-running case, as illustrated in Fig. \ref{fig:Fig1} (middle right). With an appropriate feedback ratio, the coherence established between the feedback mode and the pump is transferred to the remaining comb lines, ideally yielding a uniform linewidth across the comb and low-jitter pulse generation as shown in Fig. \ref{fig:Fig1} (bottom right). Consequently, the frequency noise PSD of the $\mu$-th comb line $S_{\Delta\nu,\mu}(\upsilon)$, is reduced proportionally with the linewidth according to $\Delta\nu_\mu=\pi S_{\Delta\nu,\mu}(\upsilon)$. In this work, we optimize the feedback ratio such that $\Delta\nu_p \approx \Delta\nu_\text{SIL}$ and use an ultra-narrow-linewidth pump laser to achieve broadband, low-noise oscillators and low timing jitter pulses. Unless otherwise specified, the laser linewidth reported in the following refers to the intrinsic linewidth derived from the flat region of the frequency-noise PSD at high offset frequencies.

\section{Experimental results}

\begin{figure*}[]
\centering
\includegraphics[width=0.975\linewidth]{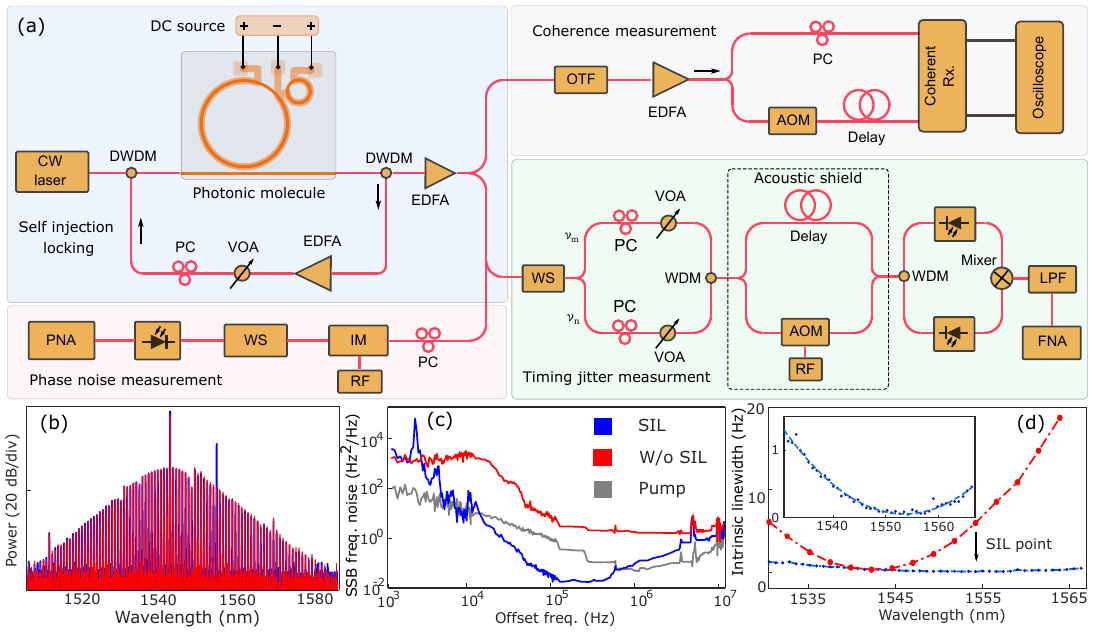}
\caption{(a) Experimental setup of the self-injection locking in the Kerr nonlinear super-efficient microcombs: self-heterodyne ultra-low-noise laser linewidth measurement (top right), electro-optic downconversion for phase noise measurement (bottom left), and fiber delay-line-based self-heterodyne timing jitter measurement on the right bottom. (b) Optical spectrum with SIL at 1553.4 nm (blue) and with no SIL (red). (c) Frequency noise spectra of the pump laser (gray) and corresponding line in a free-running state (red) and subject to optical feedback (blue). (d) Intrinsic linewidth of the super-efficient microcomb lines over the C-band without (red) and with SIL (blue). Inset: Quadratic fitting of the linewidth for SIL microcombs. OTF, optical tunable filter; AOM, acousto-optic modulator; DWDM, dense wavelength division multiplexer; IM, intensity modulator; WS, waveshaper; PNA, phase noise analyzer; FNA, frequency noise analyzer.}
\label{fig:Fig2}
\end{figure*}

In this section, we describe the operation and characteristics of the key functional blocks and discuss how optical feedback modifies the linewidth distribution of the microcomb in a manner consistent with the elastic tape model, thereby significantly enhancing coherence across the comb lines. A technical schematic of the setup used for self-injection locking and noise measurements is shown in Fig. \ref{fig:Fig2}. In an experiment, a super-efficient coupled resonator architecture \cite{helgason2023surpassing, girardi2025superefficient} with 15 million intrinsic $Q$ and $\beta_2$ = -87.8 ps$^2$/km at 1542.3 nm, measured by swept wavelength interferometry (SWI) \cite{twayana2021frequency} was used. A commercial ultra-narrow-linewidth CW laser, self-injection locked to a high-$Q$ whispering-gallery-mode microresonator near 1542.3 nm, is used as the pump laser. The microcomb is generated through thermo-optic tuning of the auxiliary and main resonances, while the pump remains fixed on the blue-detuned side of the main resonance. The generated soliton comb runs for hours with no active feedback stabilization in a standard laboratory environment. The super-efficient microcomb spectra with $35\%$ conversion efficiency without SIL (red) and with SIL (blue) is illustrated in Fig. \ref{fig:Fig2}(b).

We operate the SIL at a regime in which intrinsic spectral narrowing is insensitive to the static feedback phase as discussed in \cite{lei2024self}. The comb line at 1553.4 nm is injection locked through an external active feedback loop of length $\sim$ 53 m long with the help of dense wavelength division multiplexing (DWDM) filters at both input and output of the chip. The feedback loop has a total loss of 2.5 dB without EDFA. An EDFA is used in the loop to amplify and a polarization controller is to align the polarization of the SIL mode to TE\textsubscript{00} mode. While the feedback loop includes an amplifier, it can still operate as a passive configuration, given the high power of the SIL mode \cite{twayana2024overcoming}. The feedback loop extends the photon lifetime of the selected line. This process results in orders of magnitude reduction in the frequency noise and sub-Hertz intrinsic linewidth oscillator. The two ultra-pure anchor points lead to the dramatic improvement in the frequency noise of other modes in a way it follows the elastic tape model as discuss in the earlier section.


Here, we utilize a cross-correlated self-heterodyne technique \cite{yuan2022correlated} to measure the frequency noise of the comb lines, which allows the extraction of the ultra-low frequency noise of individual comb lines by cross-correlating signals from independent detection paths, thereby rejecting uncorrelated noise. In the schematic to the top-right Fig. \ref{fig:Fig2}(a), an optical tunable filter (OTF) selects a comb line, sends it to the Mach-Zehnder Interferometer(MZI) with an AOM and a 1 km delay line at one arm. A coherent receiver records IQ phase heterodyne beatnotes on a high-speed oscilloscope. With a cross-correlation between beatnotes, the frequency noise PSD is extracted with minimum noise of the photodetector. Figure \ref{fig:Fig2}(c) illustrates the measured single sideband (SSB) frequency noise. The SSB frequency noise of the comb line at 1553.4 nm (red) is well above the ultra-low noise pump (gray). With SIL of the comb line, the frequency noise is significantly suppressed (blue). There is an order-of-magnitude reduction in frequency noise. However, the frequency noise suppression is observed within an offset frequency of 10 kHz to 1 MHz compared to the low-phase noise pump. For a pump with higher frequency noise, two to four orders of magnitude better frequency noise is observed with the SIL over the entire Fourier frequency. At low offset frequency, frequency noise is primarily limited by phase variation due to the phase drift of the free-running feedback loop, and at high offset frequency, it is due to the limited locking bandwidth of the high-$Q$ cavity. 

The intrinsic linewidth illustrated in Fig. \ref{fig:Fig2}(d) is calculated by averaging the frequency noise PSD at an offset frequency between $3\times10^5$ and $5\times10^6$ Hz to avoid flicker noise and high offset frequency noise contribution. The measured linewidth of the pump is 0.34 Hz. With a free-running microcomb, the linewidth of comb lines follows a quadratic distribution away from the pump (red curve in Fig. \ref{fig:Fig2}(b) with measurements taken for every third comb line). Upon enabling the feedback loop and optimizing a correct feedback ratio, the reference point is shifted near the comb line selected for the feedback(blue curve in Fig. \ref{fig:Fig2}(d)).  A frequency noise of 0.025 Hz\textsuperscript{2}/Hz with SIL is demonstrated, with a corresponding linewidth of 0.058 Hz. The inset shows the quadratic distribution of the sub-Hz linewidth over the entire C-band of a microcomb under self-anchoring. To the best of our knowledge, this work presents the first experimental demonstration of a 100-GHz-spaced microcomb with sub-Hz linewidth across an extended optical bandwidth.

\section{Numerical simulation}
The linewidth distribution under SIL is studied with numerical simulations based on the Ikeda map. A comprehensive analysis of SIL dynamics under feedback is reported in \cite{lei2024self}. Here, we investigate the influence of self-injection locking (SIL) on the intrinsic linewidth distribution of microcomb lines and assess its consistency with the predictions of the elastic tape model. To elucidate the underlying physics, we adopt a single-cavity microcomb model that captures the essential dynamics of photonic molecule microcombs. Within this framework, the Raman contribution is neglected, and the analysis is restricted to pump phase noise and quantum noise. The frequency noise PSD and linewidth of the comb lines are calculated by recording the phase in multiple round-trips. With SIL disabled ($\eta=0$), the pump phase noise is equally transduced in all the comb modes. The quantum noise field coupled to the cavity is represented by a random noise amplitude\cite{paschotta2004noise} that leads to the quadratic linewidth distribution away from the pump. These are the fundamental noises in the optical frequency comb system. The self-injection locking (SIL)) technique enables overcoming this fundamental noise, which adheres to the elastic tape mode in the linewidth distribution in a similar manner to optical Kerr synchronization \cite{moille2025all}. When the SIL is enabled, the quadratic coefficient and quiet point vary for different feedback ratios.

\begin{figure}[bt]
\centering
\includegraphics[width=0.9\linewidth]{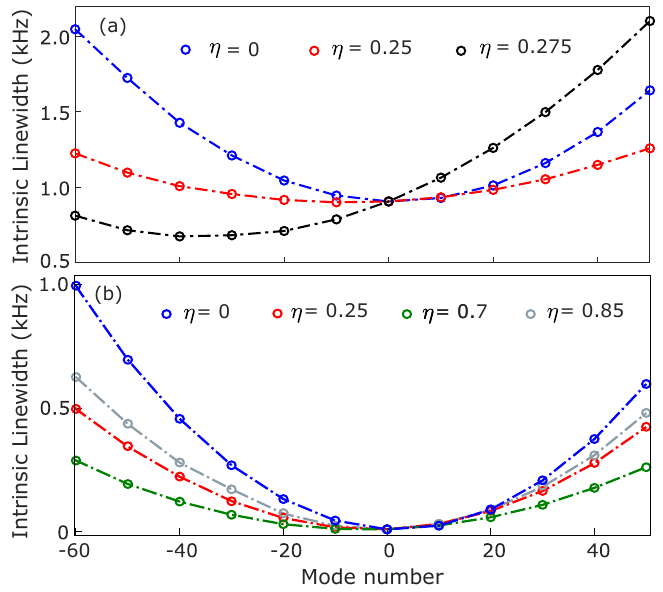}\caption{Numerical simulation of intrinsic linewidth distribution of a soliton microcomb under SIL at $\mu=-30$ for different feedback ratios. The pump, located at $\mu=0$, has an intrinsic linewidth of 910 Hz (a) and 10 Hz (b).}
\label{fig:sim}
\end{figure}

To investigate the quiet-point dynamics as a function of feedback ratio, we consider a self-injection-locked (SIL) mode at -30 and a pump intrinsic linewidth of 910 Hz. Figure \ref{fig:sim}(a) shows the intrinsic linewidth distribution for different feedback ratios, with the blue curve corresponding to the free-running case. At a feedback ratio of 0.25, the quadratic coefficient is reduced (0.11), yielding linewidths below those of all modes without SIL. This reflects suppression of phase noise, as SIL effectively anchors one comb mode and constrains global phase fluctuations. As the feedback ratio increases to 0.275, the linewidth distribution becomes asymmetric, with an increased quadratic coefficient (0.202) and a shift of the quite point to $\mu=-34$. This indicates a redistribution of phase noise away from the pump toward the new reference point, leading to enhanced noise in distant modes and increased timing jitter. These results highlight the importance of carefully optimizing the feedback ratio in SIL systems to achieve a comb spectrum with nearly uniform linewidth across all modes.

Now we investigate for a case where the frequency noise of the laser is significantly lower. Figure \ref{fig:sim}(b) is the intrinsic linewidth for different feedback ratio for the pump linewidth $\sim10$ Hz. In this regime, SIL suppresses frequency noise across all comb lines, with the quiet point remaining close to the pump. The linewidth distribution flattens with increasing feedback, as reflected by the reduction of the quadratic coefficient from 0.255 at $\eta = 0$ (blue) to 0.086 at $\eta=0.7$ (green), consistent with the elastic tape model. For stronger feedback, the distribution becomes steeper (gray), although still broader than the free-running case. A systematic exploration for the finer parameter space including feedback ratio, SIL point, and laser linewidth, is left for future work. Overall, the results indicate that SIL reshapes the noise distribution in a way that fits the elastic tape model and optimized feedback enables a substantial reduction in intrinsic linewidth across the comb, leading to reduced timing jitter. 

\section{Ultralow timing jitter}
\begin{figure*}[htb]
\centering
\includegraphics[width=\linewidth]{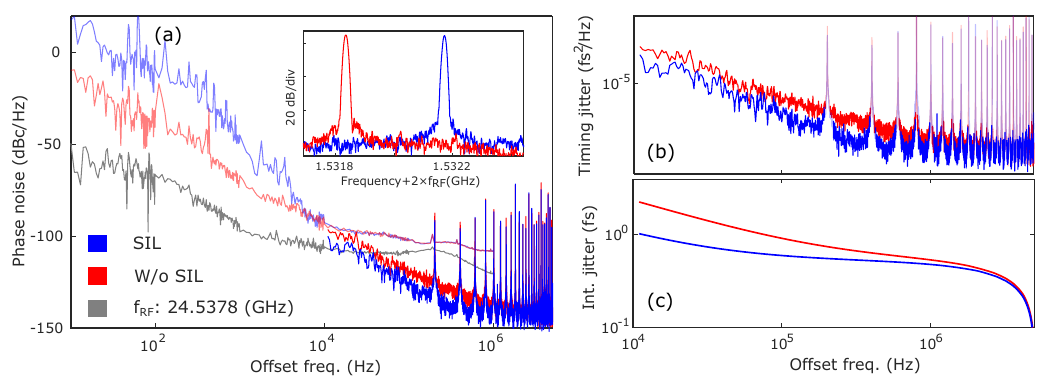}\caption{(a) SSB phase noise measured with an EO downconversion of repetition rate (translucent plots) and with a fiber delay-line-based self-heterodyne method. Inset: Downconverted repetition rate of the microcomb. (b-c) PSD timing jitter (translucent peaks arise from the delay-line-based measurement method) and integrated jitter without (red) and with (blue) SIL.}
\label{fig:Fig5}
\end{figure*}

We consider two different techniques to measure the stability of repetition rate at low and high frequency offsets. The schematic to the bottom left in Fig. \ref{fig:Fig2}(a) is a linear electro-optic (EO) downconversion of the repetition rate of the microcomb \cite{del2012hybrid}. This technique transduces the repetition rate and phase noise of the microcomb on higher-order sidebands of adjacent lines. It alleviates the stringent requirement of high-speed electro-optic devices. However, the measured phase noise has an additional penalty from the RF source noise given by $| \nu_\text{rep}\pm \Delta \nu_\text{rep}-2J(\nu_\text{RF}+\delta\nu_\text{RF}) |$ where, $J$ is the sideband or Bessel function order and $\delta\nu_\text{RF}$ is the RF frequency noise. The modulated second sideband incurs an additional 12.04 dB of phase noise on top of the RF clock's phase noise. The SSB phase noise of the rf clock (gray) and repetition rate of the comb are illustrated in Fig. \ref{fig:Fig5}(a). The measurement is limited by the phase noise of the clock beyond a 10 kHz frequency offset. We observed that the phase noise of a free-running PM microcomb (red) is orders of magnitude better than that of a single cavity microcomb reported in \cite{lei2022thermal}. Further investigations are required to solve this question. Notwithstanding, under SIL, the flicker noise of microcomb deteriorates due to the slow phase drift of the feedback loop. This is evident from the experimental result in Fig. \ref{fig:Fig5}(a) (blue) at the low frequency offset. Figure \ref{fig:Fig5}(a) in the inset shows radio-frequency beatnotes recorded on a spectrum analyzer with a resolution bandwidth (RBW) of 10 kHz. The difference in repetition rate is attributed to the change in detuning due to the SIL mode.

The high-frequency offset phase noise and timing jitter are measured using a fiber delay-based heterodyne detection method \cite{kwon2017reference}. Using this technique, a femtosecond timing jitter free-running silica microcomb is reported in \cite{jeong2020ultralow}. Figure \ref{fig:Fig2}(a) bottom-right shows the experimental setup. A programmable filter is used to filter out two comb modes near 1536 nm ($\nu_m = m\nu_\text{rep}+\nu_\text{ceo}$) and 1555 nm ($\nu_n = n\nu_\text{rep}+\nu_\text{ceo}$) to capture the frequency noise where, $\nu_\text{rep}$ is the repetition rate, $\nu_\text{ceo}$ is the carrier-envelope offset frequency, and $m, n$ are the mode numbers. In this work, a $\sim$1 km (delay time $\tau$) fiber delay self-heterodyne configuration with an AOM driven with a 55 MHz radio-frequency carrier in an acoustic shield and anti-vibration mount is used to transduce the frequency noise at $\nu_\text{aom}=55$ MHz. 

The photodetectors detect the frequency noise at each mode $\text{(}m,n\text{)}\Delta\nu_{rep}+\Delta\nu_{ceo}+\nu_{aom}$ weighted by the delay $\tau$. Here, we assume the time delay incurred on both modes is constant. The photodetected frequency noise between $\nu_m$ and $\nu_n$ at $\nu_\text{aom}$ is combined by a frequency mixer that results in frequency noise $\tau\text{(}n-m\text{)}\Delta\nu_\text{rep}$. A frequency noise analyzer (FNA) records frequency noise as a voltage noise PSD. To evaluate the corresponding phase noise PSD, the transfer function of the delay-arm interferometer is applied \cite{kwon2017reference}. This phase noise is then scaled down by the optical frequency division factor (1.9 THz) and combined with the phase noise data measured using the electro-optic (EO) downconversion technique. There is a perfect overlap in the phase noises measured near 10 KHz frequency offset (Fig. \ref{fig:Fig5}(a)). The spurious peaks at integer multiples of the fiber interferometer's FSR (1/$\tau$) arise from destructive interference of the phase at the MZI output. While the phase noise at high offset frequency is improved by enabling the SIL, the phase noise at low offset frequency is deteriorated due to the slow phase drift of an external feedback loop. We believe that an active phase stabilization of the feedback loop can improve the phase noise at the low offset frequency. Figure \ref{fig:Fig5}(b) shows the measured timing jitter PSD (at 100 GHz carrier) for a free running state (red) and subject to optical feedback (blue). The measured timing jitter PSD is decreased from $2\times10^{-6}$ fs\textsuperscript{2}/Hz to $0.28\times10^{-6}$ fs\textsuperscript{2}/Hz at 100 kHz offset frequency.  The integrated (rms) timing jitter in Fig. \ref{fig:Fig5}(c) was obtained by integration of timing jitter PSD fitted with a function $a+b/(\upsilon+ c\upsilon\textsuperscript{2})$, where $\upsilon$ stands for the offset frequency and other parameters are constant coefficients. With a SIL enabled, the integrated timing jitter decrease from 2.2 fs to 1.03 fs when integrated from 5 MHz to 11 kHz offset frequency.

\section{Discussion}
Free-running microcombs are fundamentally limited by collective spectral fluctuations arising from the unconstrained repetition rate and carrier-envelope offset degrees of freedom. Self-injection locking pins down, an otherwise drifting mode, thereby constraining the frequency noise of the comb lines without the need for self-referencing using a single driving pump laser. In this work, we validated the elastic tape model governing the linewidth distribution of a microcomb under self-injection locking (SIL). Using an ultra-low-noise pump laser, a highly efficient microcomb platform, and optimized feedback conditions, we demonstrated a power-efficient microcomb with near sub-Hz intrinsic linewidth across the entire C-band, accompanied by femtosecond-level timing jitter. While this work relies on an external feedback loop, the architecture is inherently compatible with on-chip implementation using integrated optical filtering and delay elements \cite{sun2026chip}. Such integration is less susceptible to polarization drift and slow phase fluctuations, thereby enabling significant reduction of phase noise at low offset frequencies. This advance in ultra-low-noise frequency comb generation opens a practical route toward fully stabilized microcombs on a chip-scale platform  for next generation microwave applications and high-speed sampling.

The devices demonstrated in this work were
fabricated at Myfab Chalmers. This work was supported by European Research Council (AdG GA); Vetenskapsrådet (2022-06575, 2020-00453); and the Swedish Foundation for Strategic Research (IS24-0075). F. Lei was supported by National Natural Science Foundation of China (NSFC) (Grant No. 62475042); The raw data for this work will be available from zenodo after publication.


\bibliography{ref}

\end{document}